\documentclass[12pt,preprint]{aastex}

\shorttitle{Nucleosynthesis Modes in the High Entropy Wind}
\shortauthors{Farouqi et al.}

\begin{document}


\title{Nucleosynthesis Modes in the High-Entropy-Wind of Type II Supernovae:\\ 
Comparison of Calculations with Halo-Star Observations}


\author{K. Farouqi,\altaffilmark{1,2}
K.-L. Kratz,\altaffilmark{3}    
L.I. Mashonkina,\altaffilmark{4}
B.Pfeiffer,\altaffilmark{2}
J.J Cowan,\altaffilmark{5}
F.-K. Thielemann,\altaffilmark{6}
and
J.W. Truran\altaffilmark{1,7}}


\altaffiltext{1}{
Department of Astrophysics and Astronomy, University of Chicago, 
Chicago, IL 60637, USA;
farouqi@uchicago.edu, truran@nova.uchicago.edu }
\altaffiltext{2}{
Institut f\"ur Kernchemie Universit\"at Mainz, D-55128 Mainz, Germany;
BPfeiffe@uni-mainz.de}
\altaffiltext{3}{
Max-Planck-Institut f\"ur Chemie (Otto-Hahn-Institut), D-55128 Mainz, Germany;
klkratz@uni-mainz.de}
\altaffiltext{4}{
Institute of Astronomy, Russian Academy of Science, 
RU-119017 Moscow, Russia; lima@inasan.ru}
\altaffiltext{5}{Homer L. Dodge Department of Physics and Astronomy,
University of Oklahoma, Norman, OK 73019, USA; cowan@nhn.ou.edu}
\altaffiltext{6}{
Departement Physik, 
Universit\"at Basel, CH-4056 Basel, Switzerland; F-K.Thielemann@unibas.ch}
\altaffiltext{7}{Physics Division, Argonne National Laboratory, Argonne, 
IL 60439; truran@nova.uchicago.edu }


\begin{abstract}
While the high-entropy wind (HEW) of Type II supernovae remains
one of the more promising sites for the rapid neutron-capture ($r$-)
process, hydrodynamic simulations have yet to reproduce 
the astrophysical conditions under which the latter occurs. 
We have performed large-scale network calculations within an extended parameter
range of the HEW, seeking to identify or to constrain the necessary conditions 
for a full reproduction of all $r$-process residuals 
N$_{r,\odot}$=N$_{\odot}$-N$_{s,\odot}$ by comparing the results with recent 
astronomical observations. 
A superposition of weighted entropy trajectories results in an excellent 
reproduction of the overall N$_{r,\odot}$-pattern
beyond Sn. For the lighter elements, from the Fe-group 
via Sr-Y-Zr to Ag, our HEW calculations indicate a transition from the need for 
clearly different sources (conditions/sites) to a possible co-production with 
$r$-process elements, provided that a range of entropies are contributing.
This explains recent halo-star observations of a clear non-correlation of Zn 
and Ge and a weak correlation of Sr - Zr with heavier $r$-process elements. 
Moreover, new observational data on Ru and Pd seem to confirm also a partial
correlation with Sr as well as the main $r$-process elements (e.g. Eu).
\end{abstract}


\keywords{nuclear reactions, nucleosynthesis, abundances --- 
stars: abundances ---  
stars: Population II}


\section{INTRODUCTION}
%\label{} 
A rapid neutron-capture process ({\it r}-process) is traditionally believed 
to be responsible for the synthesis of about half of the heavy elements above 
Fe \citep{b2fh,cam}. The astrophysical site in which this mechanism operates 
is, however, still uncertain. For this reason, a model-independent approach, 
i.e. the 
classical ``waiting-point" approximation, has been utilized for many years 
(see, {\it e.g.}, 
\citet{cow91,fk2l,kr07a}). This simple model has helped to 
gain increased insight into the systematics of the $r$-process: {\it e.g.} its 
dependence on nuclear-physics input and its sensitivity to astrophysical 
conditions (see, {\it e.g.}, \citet{pf01,kr07b}). In realistic explosive 
scenarios, the necessary conditions for high neutron-to-seed ratios 
(Y$_n$/Y$_{seed}$) can only be obtained in very neutron-rich low-entropy (S) 
environments,  related {\it e.g.}, 
to neutron-star ejecta from neutron-star mergers (NSM),  or in moderately 
neutron-rich high-S scenarios, such as the high-entropy wind (HEW) of core 
collapse (Type II) supernovae (SNe II). As observations of heavy element 
abundance patterns in metal-poor stars in the early Galaxy 
(see, {\it e.g.}, \citet{snco06,cowsne06})
and galactic chemical-evolution considerations both seem to disfavor 
NSM \citep{argast}, we will focus here on the HEW scenario. Moreover, since 
even the most recent hydrodynamical simulations encounter problems in the 
time-evolution of the HEW bubble and/or in the attainment of sufficiently 
high entropies, we continue to use parameterized dynamic network calculations 
to explore the dependence on nuclear properties and highlight a detailed 
understanding of the HEW nucleosynthesis 
processing \citep{far08a,far08b}.

\section{CALCULATIONS AND RESULTS}
% \label{}
The concept of a HEW arises from considerations of the 
newly born proto-neutron star in core-collapse supernovae. In this scenario, 
the late neutrinos interact with matter of the outermost neutron-star layers, 
leading to moderately neutron-rich ejecta with high entropies 
(see, {\it e.g.}, 
\citet{1,2,3,7a,7b,4,5,6}). In the present calculations, we follow 
the description of adiabatically expanding mass zones as previously 
utilized in \citet{4}. The nucleosynthesis calculations up to charged-particle 
freeze-out were performed with the latest Basel code (but without including 
neutrino-nucleon/nucleus interactions). The reaction rates were calculated by 
means of the statistical-model program NON-SMOKER \citep{9}. The $r$-process 
network code now contains  updated experimental and theoretical nuclear 
physics input on masses and $\beta$-decay properties, as  outlined 
in \citet{kr07b} and used in our earlier studies within the site-independent 
``waiting-point" approximation \citep{kr07a}. \\ 

After charged-particle freeze-out, the expanding, and eventually ejected, mass 
zones have different initial entropies (S${\sim}$T$^3$/$\rho$ [k$_b$/Baryon]), 
so that the overall explosion represents a superposition of entropies. The 
ratio of free neutrons to ``seed'' nuclei (Y$_n$/Y$_{seed}$) is a function of 
entropy and, for high S, yields rapid neutron captures which can form 
the heaviest $r$-process nuclei. Furthermore, the ratio 
Y$_n$/Y$_{seed}$ is correlated to the three main parameters of the HEW, i.e. 
the electron abundance (Y$_e$=Z/A), the entropy (S) and the expansion velocity 
(V$_{exp}$) \citep{7b,4}.  We determined the simple expression 
Y$_n$/Y$_{seed}$ = $ 10^{-11}$$\cdot$V$_{exp}$ (S/Y$_e$)$^3$, valid in the 
parameter ranges 0.4$<$Y$_e$$<$0.495, 1500$<$V$_{exp}$$<$30000 and 1$<$S$<$350.
This ratio Y$_n$/Y$_{seed}$ provides a measure of the strength 
of the $r$-process. 

In the classical $r$-process approach, a range of neutron densities 
(10$^{20}$$<$n$_n$$<$10$^{28}$) is necessary to reproduce the full 
distribution of the $r$-process ``residuals" 
(N$_{r,\odot}$=N$_\odot$-N$_{s,\odot}$; \citet{kaep}) up to the Th, U 
region \citep{fk2l,kr07b,kr07a}. In the HEW, at a given Y$_e$ 
and V$_{exp}$, the entropy (or the correlated Y$_n$/Y$_{seed}$ ratio) will 
play this role. In the following, we will present selected nucleosynthesis 
results as a function of entropy (or Y$_n$/Y$_{seed}$) for the  
(realistic and astrophysically interesting) 
choices Y$_e$=0.45 and V$_{exp}$=7500 km/s 
- values taken from the much larger parameter space which was analyzed - 
and compare our HEW predictions 
to recent astronomical observations. 

We first want to identify the abundance distributions that can be produced 
when utilizing the above parameter combination. In Fig.~1,
we show the abundance correlations of $\alpha$-particles (Y$_\alpha$), 
heavy ``seed" nuclei (Y$_{seed}$) and free neutrons (Y$_n$) as functions of 
entropy at the freeze out of charged-particle reactions. Over the entire 
entropy range displayed in this figure, most of the matter is locked 
into $\alpha$-particles. In contrast to the smooth trend 
of Y$_{\alpha}$, the abundance distributions of Y$_{seed}$ and Y$_n$ are 
strongly varying with entropy. For different values of Y$_e$ and V$_{exp}$ this behavior is shifted
to lower/higher entropies. From the Y$_{seed}$ and Y$_n$ slopes it becomes 
evident that the HEW predicts -- at least -- two clearly different 
nucleosynthesis modes. 

For the low entropy region (1$\le$S$\le$110), the concentration 
of free neutrons is negligible; hence, the nucleosynthesis in this region 
is definitely not a classical neutron-capture process but rather a 
charged-particle ($\alpha$-) process. For higher entropies
Y$_n$/Y$_{seed}$ ratios are increasing smoothly, resulting for the region 
110$\le$S$\le$150 in a neutron-capture component which resembles a 
classical {\it ``weak''} $r$-process. 
For even higher entropies (150$<$S$<$300),  
enough free neutrons are available to yield 
a classical {\it ``main''} $r$-process \citep{7b,meyer97,4,pf01}.  
In Table 1 we show the contributions of certain entropy ranges to the 
production of elements (Z) in percent, under the assumption of equal mass 
contributions per entropy interval for 1$<$S$<$300.

\begin{itemize}

\item{}
As can be seen from Table 1, in the lowest entropy 
range (1$<$S$<$50), we obtain a ``normal" $\alpha$-rich freeze out, mainly 
producing stable or near stable isotopes of elements in the region Fe to Sr
(for further details, see Table 1 in \citet{far08b}). It should be noted 
that for varying choices of Y$_e$=0.45-0.49, 
the classical {\it ``s-only"} isotopes up to $^{96}$Mo or even
light {\it ``p}-nuclei" up to $^{106}$Cd can be produced. 

\item{}
In the next higher entropy range (50$<$S$<$110; column 2 of Table 1), we 
again find that there are not enough free neutrons available to effect 
a neutron-capture process. Under these entropy conditions, however, the 
seed composition at freeze out is already shifted to the neutron-rich side 
of $\beta$-stability, including $\beta$-delayed neutron ($\beta$dn) 
precursor isotopes in the 80$<$A$<$100 mass region \citep{26}. 
The resulting Y$_{\beta dn}$/Y$_{seed}$ conditions provide a low 
neutron-density, s-like environment. 

\item{}
In the subsequent entropy range (110$<$S$<$150), the density of free neutrons 
(1$\le$Y$_n$/Y$_{seed}\le$10) becomes high enough to start a {\it ``weak''} 
$r$-process up to the rising wing of the A$\simeq$130 N$_{r,\odot}$ peak.
%but not including significant production of $^{129}$I/Xe. 
As shown in the 3$^{ rd}$ column of Table 1, under these conditions 
substantial concentrations of the elements Ru to Ag are produced. 

\item{}
For high entropies (150$<$S$<$300; now with 13$\le$Y$_n$/Y$_{seed}\le$155) 
the HEW predicts a very robust {\it ``main''} $r$-process, starting with 
the N=82 $r$-process progenitor isotopes of Tc to Rh at the onset of the 
A$\simeq$130 peak and reaching up to the Th, U actinide region. 

\end{itemize}

As a function of time, the HEW will eject matter with varying values of S, 
Y$_e$ and V$_{exp}$. If one assumes that equal amounts of ejected material 
per entropy interval are contributing, the sum of the abundance contributions 
is weighted according to the resulting Y$_{seed}$ as a function of entropy.
Such a choice yields a SS-like isotopic abundance distribution 
(N$_{r,calc}$) as displayed in Fig.~\ref{figure2} in comparison to
the standard $r$-process ``residuals'' N$_{r,\odot}$. 
The N$_{r,calc}$ distribution represents a superposition 
of 15 equidistant S-components in the range 160$\le$S$\le$287.
The lower limit of S=160 has been chosen to restrict the calculations to 
representative ``main'' $r$-process conditions. 
Consequently, for this parameter choice no ``best fit'' to the light region 
below the A$\simeq$130 peak is anticipated. The upper limit of S=287 
has been chosen to ensure that fission recycling remains negligible. 
We note that excellent overall agreement of the N$_{r,calc}$ 
distribution with the observed N$_{r,\odot}$ pattern is attained from the 
rising wing of the A$\simeq$130 peak up to the Pb, Bi 
``spike''. 

It is also obvious that this superposition choice is not reproducing abundances
from the Fe-group to the rising wing of the  A$\simeq$130 peak. There have been
a number of suggestions to fill in this region with a multiplicity of 
nucleosynthesis processes. As such elements are apparently already existing at 
low metallicities, but are not explainable by the traditional 
metallicity-dependent (secondary) $s$-process, a LEPP (light element primary)
process was invoked by \citet{14}, initially related to s-like neutron captures.
\citet{15}, following the initial argument of \citet{7a,7b}, consider 
these elements to be primarily produced due to charged particle reactions (CPR).

The HEW approach, with different choices of entropy superpositions for
S$<$110 is such a charged particle process, for 110$<$S$<$150 it also 
results in small
neutron densities. We compare these predictions with recent astronomical 
observations of the abundances of elements between Cu and Ag, covering 
the range from ``$r$-process poor" stars like \object{HD 122563}
up to ``$r$-process rich" stars like \object{CS 22892-052}
(see, {\it e.g.}, 
Fig.~3 in \citet{far08a}). A crucial test would be that at least within a
narrow S-range, responsible for neighboring nuclei, the observed
element ratios should be reproduced. Such a test avoids uncertainties in the
choice of realistic entropy superpositions. When doing so for the Zn/Ge ratio,
where observations \citep{23} show a factor beyond 100,
our calculations - producing these nuclei for
S$<$ 50 - would predict a ratio of 5. Thus, the present HEW 
conditions for Y$_e$-values $<0.5$ disagree by a factor of 25.
This disagreement can be avoided, when permitting proton-rich environments
as discussed in the $\nu$p-process \citep{25a,25b} during the 
very early phases of the neutrino wind, when even proton-rich conditions with
Y$_e$$>$0.5 are obtained.
Such conditions are expected in every core collapse
supernova. The conclusion which emerges here, is that although both 
environments involve charged-particle processes, the Y$_e$ dependence plays a 
crucial role. 
An alternative is a strong primary $s$-process, occurring for massive 
stars at very low metallicities \citep{pignat08}. 

In Fig.~3 we plot the LEPP abundance ratios, log(X/Zr) as a function of 
atomic number in the range 29$\le$Z$\le$50 and compare the 
observational data with the predictions from two different nucleosynthesis 
approaches: (i) the present HEW superpositions with weights taken as in Table 1
and (ii) the classical ``waiting-point'' 
model for a range of n$_n$-conditions which fit best the low mass region for
$r$-process residuals (permitting also an extended seed composition below Fe).
The observed log(X/Zr) pattern shows a decreasing slope with atomic number
with a pronounced odd-even Z staggering. From Sr (Z=38) 
upwards, this trend is best reproduced by the HEW (with equal mass 
superpositions in entropy S). In contrast, the high abundances observed for 
Cu (Z=29) and Zn (Z=30) obviously cannot be reproduced by any of the 
models, while proton-rich environments and the $\nu$p-process \citep{25a,25b}
show the option to do so. At the moment, we only can conclude  
that the major abundance fractions of the elements with Z$<$38 are {\it not} 
produced together with the Z$\ge$38 elements. Therefore, 
these light trans-Fe elements are also
not produced under the same nucleosynthetic conditions as Eu, and should 
be uncorrelated with the ``main'' $r$-process. 
This result is, indeed, 
confirmed for Ge abundances recently obtained from Hubble Space Telescope 
observations \citep{23}.     

We now consider the three LEPP neighbors Sr, Y and Zr, 
for which two independent large data sets from (i)  \citet{bark} and 
 \citet{mash} (B\&M set) and (ii) from 
 \cite{francois} exist. These data sets agree very well in
showing that the abundances of Sr, Y and Zr, viewed as separate elements, are
not tightly (but partially) correlated with the (almost pure) ``main'' 
$r$-process element Eu. 
The elemental abundance ratios of log(Sr/Y/Zr) from Galactic 
halo stars, {\it e.g.} as a function of metallicity [Fe/H], $r$-process enrichment
[Eu/H] or LEPP enrichment [Sr/H] have been shown to exhibit a robust, 
constant pattern (see, {\it e.g.}, \citet{14,15,mash,francois}). 
Early studies \citep{7a} seemed to indicate a strong $Y_e$ dependence for the
light trans-Fe elements (for entropies S$<$50).
We have done calculations to predict the elemental ratios in this
mass region for a variety of entropies. What is noticed is, that for the higher
entropies which contribute to this mass range (60$<$S$<$110, see Table 1)
the elemental ratios are converging to the observed ones, as indicated for 
Sr/Zr in the top frame of Fig.~4, but has also been tested for Sr/Y and Y/Zr. 
For lower entropies, also contributing to these elements, we see for each Y$_e$
a variation as a function of entropy, leading on average also to values close
to the observed ones. Thus, when integrated over the relevant S-conditions,
these abundances seem to be independent of Y$_e$ and reproduce the 
observed abundance ratios very well. This indicates that in a  
superposition over entropies the Sr/Y/Zr elements can be coproduced with the
observed abundance ratios (see in Fig.~4 - lower two panels - (Sr/Y,Zr) as a 
function of ``$r$-process enrichment" in the range $-3.5\le$[Eu/H]$\le-0.5$
from the B\&M data set and also data from 
\citet{24,francois,25,18a,18b,18,ivans,lai,sne03}, but 
also also Fig.3 for Z=38, 39 and 40).

The mean values of the observed abundance ratios  
(log(Sr/Y) = 0.83 $\pm$ 0.15 and log(Sr/Zr)  =  0.09 $\pm$ 0.18)
are very well reproduced (log(Sr/Y)=0.83 and log(Sr/Zr)=0.13)  by the low 
entropy (S$<$110) charged-particle ($\alpha$-) component of the HEW, dominating 
that mass range. When comparing the observed data set to HEW predictions, 
restricted to higher entropies (110$\le$S$\le$280) which lead to a 
neutron-capture $r$-process, the resulting ratios log(Sr/Y)=0.99 and
log(Sr/Zr)=$-1.38$) indicate that this nucleosynthesis 
mode is unimportant for that mass range. This is consistent with the fact 
that the element correlations [Zr/Fe] vs. [Eu/Fe] (in analogy to Fig.6 in 
\citet{23}) clearly show - now for an ensemble of 83 halo stars 
in the range -0.7$\le$[Eu/Fe]$\le$1.7 - that the above LEPP elements are not 
tightly correlated with the $r$-process indicator Eu.

We also want to call attention to the recent accumulation of abundance data 
on the two light platinum-group elements (PGE) Ru (Z=44) and Pd (Z=46) for the 
extremes of ``r-poor" stars (such as HD 122563; \citet{18b}) 
and ``r-rich" stars (such as CS 22892-052; \citet{sne03}). 
Table 1 indicates that both elements can be
co-produced in substantial fractions by the charged-particle component (i.e.
alpha-freeze-out, together with Sr-Zr) and the neutron capture (r-) component 
together with the nucleosynthesis of A=130 peak elements Sn-Te.
In halo stars and r-rich but alpha-component-poor stars, the mean values of 
the observations are log(Pd/Sr)=-0.95 and log(Pd/Eu)=0.81. For r-poor but 
alpha-component-rich stars the observed abundance ratios are log(Pd/Sr)=-1.16 
and log(Pd/Eu)=+1.5. This indicates that for the halo and r-rich
stars the dominant r-component of Pd correlates with Eu, whereas for the
r-poor stars the alpha-component of Pd dominates and is correlated with Sr. 

The question remains whether the observed Ru/Pd ratios can also be 
reproduced by our HEW superpositions. Fig.3 indicates a good average trend
of the canonical equal mass superposition per entropy for Z=44 and 46.
Observations in low metallicity stars are scarce, log(Ru/Pd) seems to cluster 
around 0.27$\pm$0.05 for normal halo stars (from determinations in this study 
and in \citet{ivans}) and around 0.39 for two r-rich stars \citep{25,sne03}. 
On the other hand, two stars which show features of the weak $r$-process are 
reported to have log(Ru/Pd)=0.47 \citep{18,18b}, although it is worth noting 
that the Pd abundance in these stars was derived from the only Pd I 3404 line.
%there are arguments
%that the uncertainties in oscillator strengths could lower this value.
If we analyze the predictions obtained from entropy components
contributing to the mass region of interest, we find a downshift in log(Ru/Pd)
by about 0.2 between the entropy regions 100$<$S$<$150 and 
150$<$S$<$200, i.e. lower Ru/Pd ratios when changing from 
weak to main $r$-process sources, which is consistent with observations.
However, the absolute value is Y$_e$-dependent (with a ratio close to 0.5
rather than 0.39 for Y$_e$=0.45 and high entropies). 
Thus, if the nuclear input is sufficiently
reliable, future improvements in models and observations could provide 
further insight into the necessary features of the astrophysical site 
(including Y$_e$-requirements)§.

\section{SUMMARY AND CONCLUSIONS} 

We have performed large-scale dynamical network calculations in the context 
of an adiabatically expanding high-entropy wind as expected in core-collapse 
SNe II. 
We find that the correlated parameters that act to determine the strength 
of the astrophysical $r$-process, reflected in the ratio 
Y$_n$/Y$_{seed}$ =  $ 10^{-11}$$\cdot$V$_{exp}$ (S/Y$_e$)$^3$, 
show a surprisingly robust picture for 
the production of heavy elements beyond Fe. 
Our results suggest astrophysical conditions for the HEW scenario 
which reproduce the solar r-abundances quite satisfactorily without invoking
more exotic nucleosynthesis scenarios or nuclear physics assumptions.

We also tested how important the HEW scenario is for the light-Z region 
between Cu and Rb. Neither the Cu/Zn/Ge abundance ratios nor the absolute 
yields observed in halo stars can be reproduced in an entropy 
superposition with Y$_e$$<$0.5, which clearly indicates another
nucleosynthesis origin. For Sr to about Mo the dominating production is related
to low entropies (S$\le$110), where the charged-particle ($\alpha$-rich) 
freeze-out results in entropy 
dependent ``seed'' nuclei smoothly shifting from $\beta$-stability 
to the neutron-rich side. However, the entropy conditions relevant for the
respective elements lead to element ratios, e.g. Sr/Y/Zr or Ru/Pd which seem
consistent with astronomical observations.
Rapid neutron-capture nucleosynthesis only sets in at higher entropies, 
where the range 110$\le$S$\le$150 produces elements up to the rising wing 
of the A$\simeq$130 N$_{r,\odot}$ peak under   ``weak''
$r$-process conditions, whereas for 150$\le$S$\le$300 a robust ``main'' 
$r$-process between A$\simeq$120 and the actinide region occurs. The HEW 
with an entropy superposition of equal mass ejecta per entropy interval
is able to reproduce the overall 
N$_{r,\odot}$ ``residuals'' beyond Sn, as well as all major recent 
observations from metal-poor halo stars. 

Thus, all heavy elements beyond Sr and their classical r-(residual) abundances 
can potentially be reproduced in entropy superpositions, which for S$\ge$60 
seem essentially independent of Y$_e$ (in the range 
0.45$\le$Y$_e\le$0.49). In the lower entropy range results are Y$_e$-dependent,
and in addition to $r$-nuclei also classical ``$s$-only'' isotopes as well 
as $p$-nuclei can be produced in the range $^{54}$Fe up to $^{106}$Cd. 
These results provide the means to 
substantially revise the abundance estimates of different primary 
nucleosynthesis processes for elements in the 
historical ``weak-$s$''/``weak-$r$'' 
process region  and to quantify their correlation with 
the ``main'' $r$-process. 

To obtain more quantitative answers to questions concerning the 
astrophysical site of the compositions of the LEPP elements between 
Sr (Z=38) and Cd (Z=48), as well as all of the $n$-capture elements,
will require more and higher quality observational 
data and also more realistic values of entropy superpositions 
derived from hydrodynamical models.

We thank R. Gallino for helpful discussions. 
Partial financial support for this research was provided  
by the Deutsche
Forschungsgemeinschaft (DFG), 
the Helmholtz Gemeinschaft, 
the NSF, the DOE as well as the Swiss NSF.

\newpage
\begin{table}[t]
 \setlength{\tabcolsep}{4mm} 
\caption{HEW-contributions to the production 
of selected elements (elemental abundance, Y(Z) in $\%$) 
in the Fe to Ag region for different entropy ranges. 
Column (1): S-range 1$<$S$<$50, normal $\alpha$-freezeout; column (2): S-range 
50$<$S$<$110, neutron-rich $\alpha$-freezeout 
with $\beta$dn-recapture; column (3): 
S-range 110$<$S$<$150, {\it ``weak''} 
$r$-process; column (4): S-range 150$<$S$<$300, 
{\it "main"} $r$-process. For discussion, see text.}
\begin{center}
\begin{tabular}{ccccc}
\hline
 Z & 1 & 2 & 3 &  4 \\
\hline
  26  &100.00  &  0.00  &  0.00  &  0.00\\
  30  & 99.2   &  0.5   &  0.3   &  0.00\\
  32  & 85.3   &  4.7   &  8.8   &  1.2 \\
%  36  & 23.17  & 72.68  &  2.93  &  1.21\\
  38  & 79.9   & 18.3   &  1.4   &  0.3 \\
  39  & 61.4  & 37.4  &  0.9  &  0.3 \\
  40  & 14.0   & 81.0  &  4.7  &  0.3 \\
  42  &  0.03  & 63.6  & 31.7  &  4.7\\
  44  &  0.00  & 27.4  & 61.4  & 11.2\\
  46  &  0.00  & 11.5  & 66.9  & 21.6\\
  47  &  0.00  &  3.7  & 71.3  & 25.0\\
\hline
\end{tabular}
\end{center}
\end{table}

%% the file, uncomment the \plotone, \plottwo, and
%% If you need a layout that cannot be achieved with \plotone or
%% \plottwo, you can invoke the graphicx package directly with the
%% \includegraphics command or use \plotfiddle. For more information,
%% please see the tutorial on "Using Electronic Art with AASTeX" in the
%% documentation section at the AASTeX Web site,
%% http://www.journals.uchicago.edu/AAS/AASTeX.
%%
%% The examples below also include sample markup for submission of
%% supplemental electronic materials. As always, be sure to check
%% the instructions to authors for the journal you are submitting to
%% for specific submissions guidelines as they vary from

%% This example uses \plotone to include an EPS file scaled to
%% 80% of its natural size with \epsscale. Its caption
%% has been written to indicate that additional figure parts will be
%% available in the electronic journal.

\newpage
\begin{figure}
%\plotone{new6.ps}
\epsscale{0.90}
\plotone{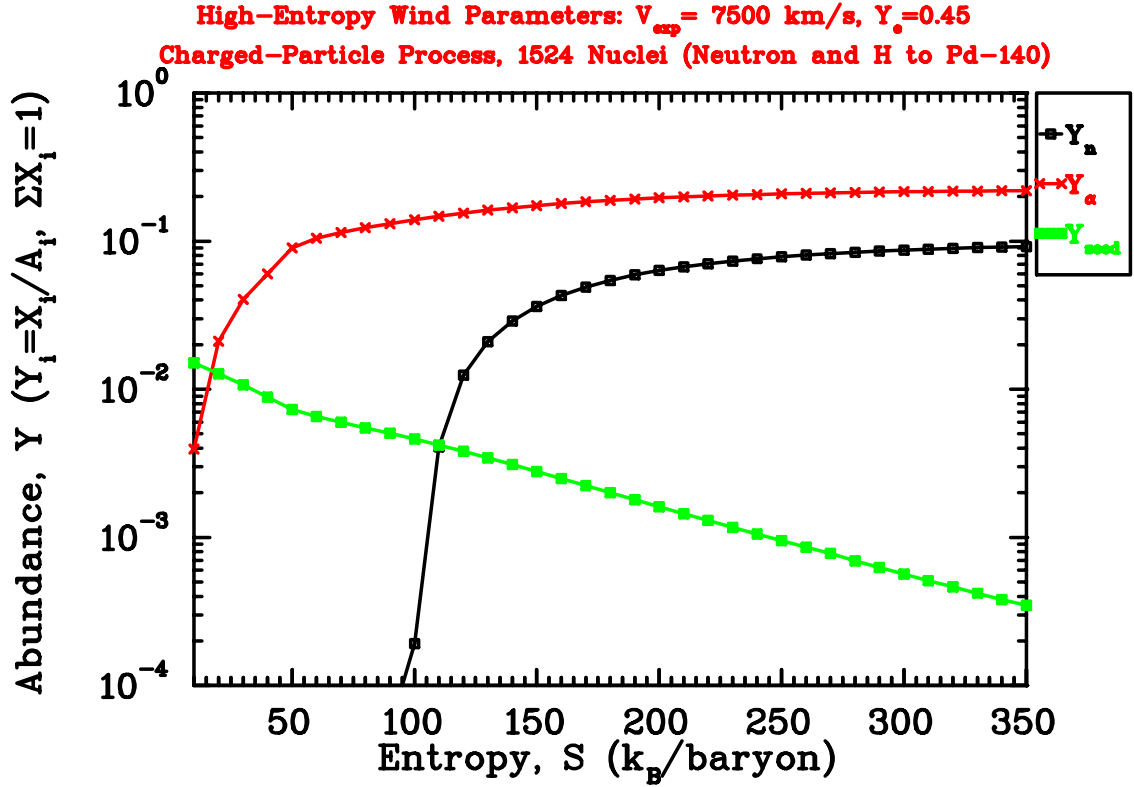}
%\vskip -0.8truein
\caption{
  Abundance distributions (Y$_i$=X$_i$/A$_i$, $\Sigma$X$_i$=1) of $\alpha$-particles, 
   heavy ``seed'' nuclei and free neutrons as function of entropy S at the 
   freezeout of charged-particle reactions at T$_9$ $\simeq$ 3.
\label{figure1}}
\end{figure}

\newpage
\begin{figure}
\epsscale{0.90}
\plotone{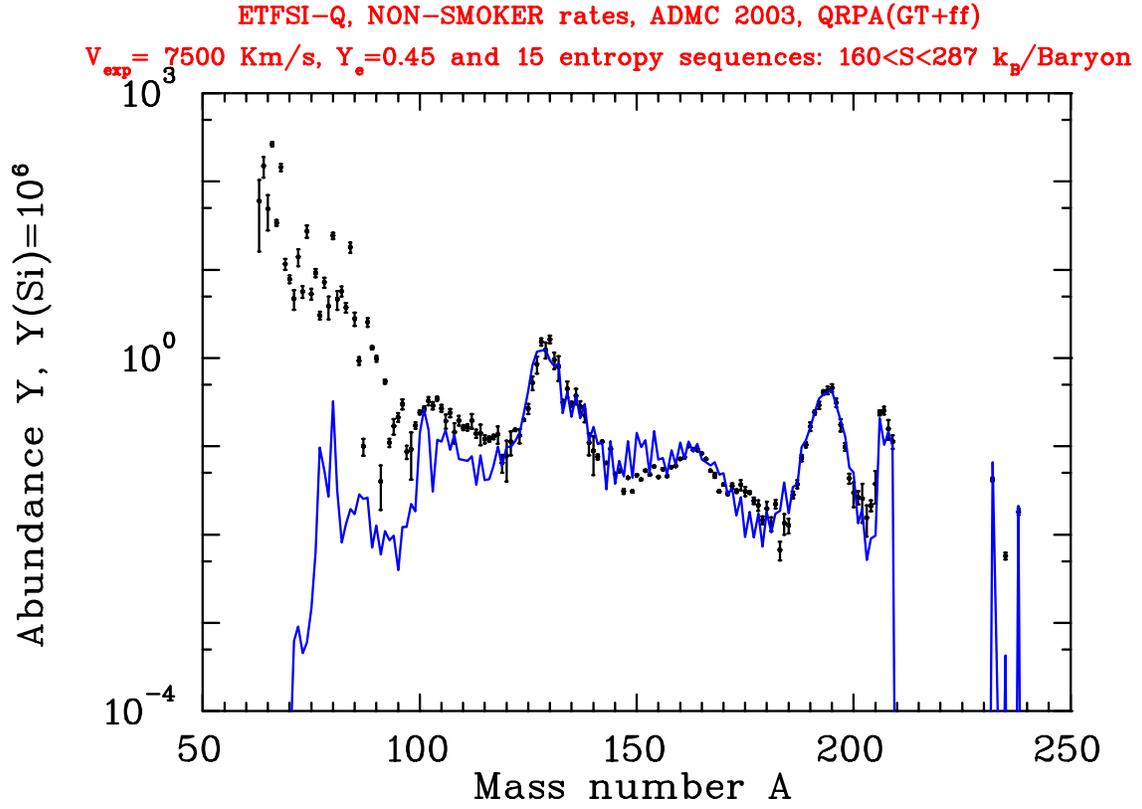}
\caption{Comparison of the N$_{r,\odot}$ distribution 
(data points \citep{kaep}) 
with predicted isotopic abundances (full line) from a weighted superposition 
of 15 HEW entropy components in the range 160$\le$S$\le$287. For further 
details and discussion, see text.} 
\label{figure2}
\end{figure}

\newpage
\begin{figure}
\epsscale{0.90}
\plotone{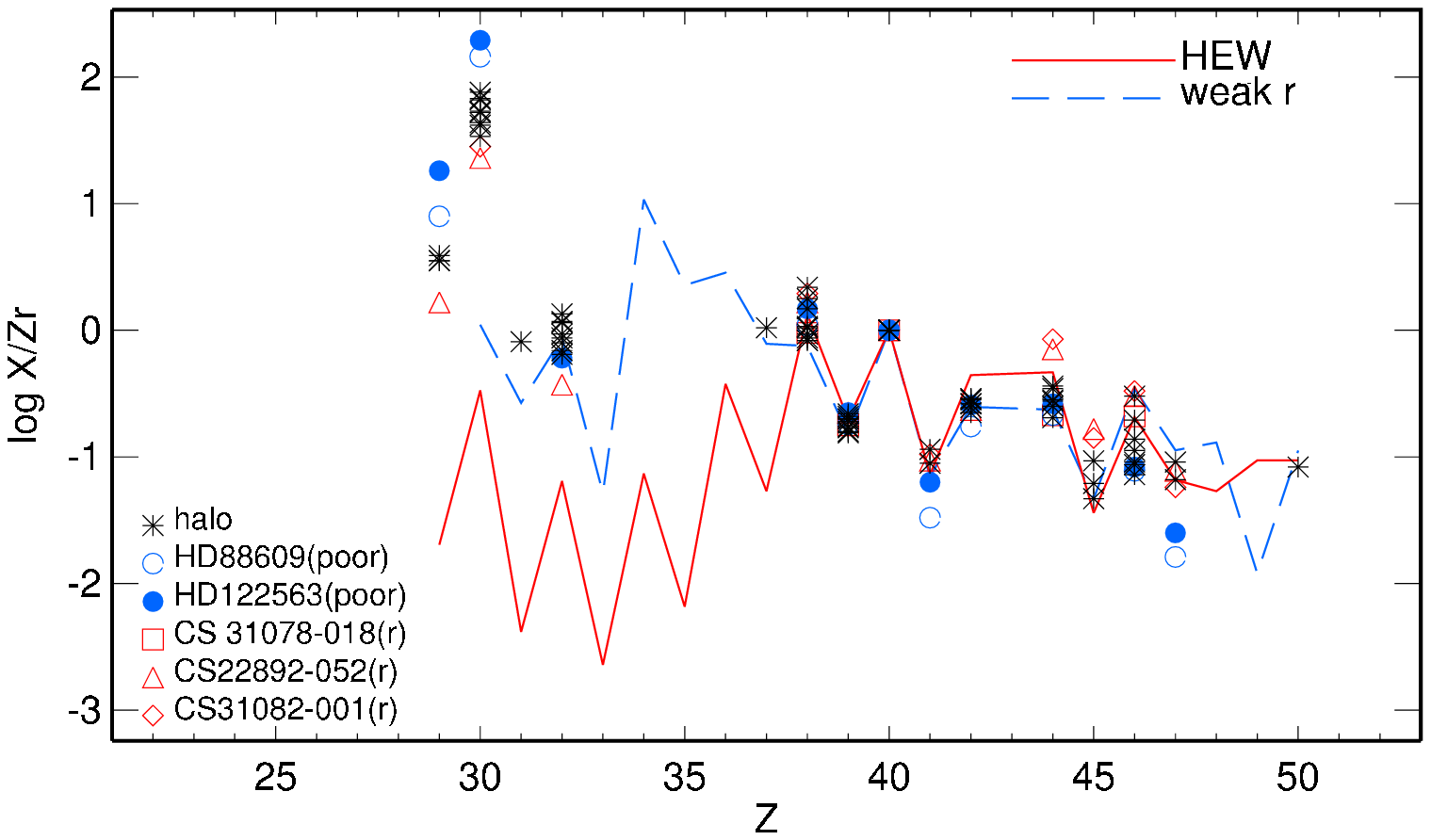}
\caption{
Elemental abundance ratios of log(X/Zr) as a function of atomic number Z. 
The observational data in the 
LEPP region between Cu (Z=29) and Ag (Z=47) for selected halo stars are 
given by different symbols 
explained in the lower left corner of the figure. These 
data are compared with predictions from two different nucleosynthesis 
approaches; (i) the present HEW model  
(full line), (ii) 
the classical ``waiting-point" model for a ``weak'' $r$-process with a low
n$_n$-range a solar-like Si -- Cr seed composition 
(dashed line). 
For discussion, see text. 
} 
\label{figure3}
\end{figure}

\newpage
\begin{figure}
\epsscale{0.80}
\plotone{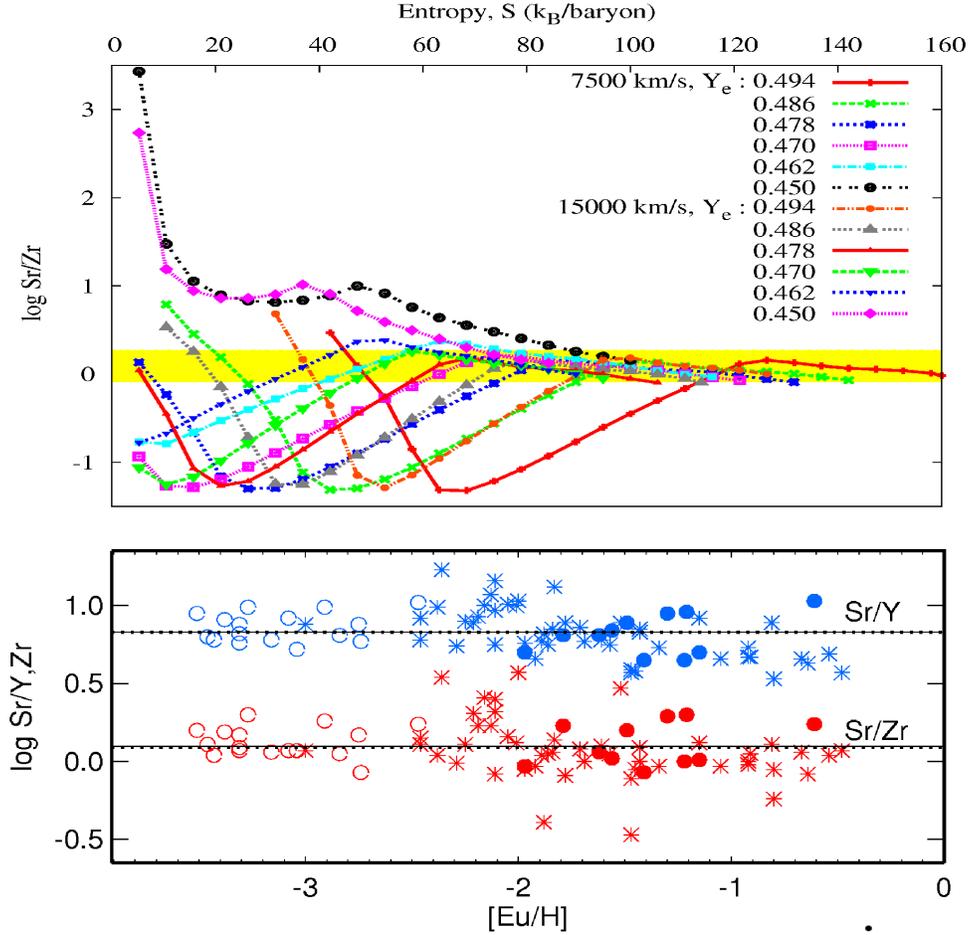}
\vskip -0.2in
\caption{
Top panel: Sr/Zr ratio as predicted from HEW 
calculations for different values of Y$_e$ as a function of 
entropy S. The observed value is shown by gray color band. Bottom panels:
Elemental abundance ratios of log(Sr/Y,Zr) as a function of $r$-process 
enrichment [Eu/H] 
for halo stars in our Galaxy. The observational data 
%of \citet{23,bark,mash,18,ivans} (
with their mean values shown as full lines are compared to the abundance 
predictions of our HEW model (dotted lines, partly indistinguishable 
from the full lines) for the low-S 
%range (1$\le$S$\le$110) of the $\alpha$-component. 
range (1$\le$S$\le$110) of the $\alpha$-component.
For discussion and the sources of observational data, see text. 
}
\label{figure4}
\end{figure}

\end{document}